\documentclass[aps,prl,twocolumn,showpacs,superscriptaddress]{revtex4}
\usepackage{graphicx}
\usepackage{color}

\begin{document}
\title{Strongly anisotropic spin-orbit splitting in a two-dimensional electron gas}
\author{Matteo Michiardi}
\affiliation{Department of Physics and Astronomy, Interdisciplinary Nanoscience Center (iNANO), Aarhus University, 8000 Aarhus C, Denmark}
\author{Marco Bianchi}
\affiliation{Department of Physics and Astronomy, Interdisciplinary Nanoscience Center (iNANO), Aarhus University, 8000 Aarhus C, Denmark}
\author{Maciej Dendzik}
\affiliation{Department of Physics and Astronomy, Interdisciplinary Nanoscience Center (iNANO), Aarhus University, 8000 Aarhus C, Denmark}
\author{Jill Miwa}
\affiliation{Department of Physics and Astronomy, Interdisciplinary Nanoscience Center (iNANO), Aarhus University, 8000 Aarhus C, Denmark}
\author{Moritz Hoesch}
\affiliation{Diamond Light Source, Harwell Campus, Didcot, OX11 0DE, United Kingdom}
\author{Timur K. Kim}
\affiliation{Diamond Light Source, Harwell Campus, Didcot, OX11 0DE, United Kingdom}
\author{Peter Matzen}
\affiliation{Department of Physics and Astronomy, Interdisciplinary Nanoscience Center (iNANO), Aarhus University, 8000 Aarhus C, Denmark}
\author{Jianli Mi}
\affiliation{Center for Materials Crystallography, Department of Chemistry, Interdisciplinary Nanoscience Center (iNANO), Aarhus University,
8000 Aarhus C, Denmark}
\affiliation{Institute for Advanced Materials, Jiangsu University, Zhenjiang 212013, China}
\author{Martin Bremholm}
\author{Bo Brummerstedt Iversen}
\affiliation{Center for Materials Crystallography, Department of Chemistry, Interdisciplinary Nanoscience Center (iNANO), Aarhus University,
8000 Aarhus C, Denmark}
\author{Philip Hofmann}
\affiliation{Department of Physics and Astronomy, Interdisciplinary Nanoscience Center (iNANO), Aarhus University, 8000 Aarhus C, Denmark}
\email[]{philip@phys.au.dk}

\date{\today}
\begin{abstract}
Near-surface two-dimensional electron gases on the topological insulator Bi$_2$Te$_2$Se are induced by electron doping and studied by angle-resolved photoemission spectroscopy. A pronounced spin-orbit splitting is observed for these states. The $k$-dependent splitting is strongly anisotropic to a degree where a large splitting ($\approx 0.06$~\AA$^{-1}$) can be found in the $\bar{\Gamma}\bar{M}$ direction while the states are hardly split along $\bar{\Gamma}\bar{K}$. The direction of the anisotropy is found to be qualitatively inconsistent with results expected for a third-order anisotropic Rashba Hamiltonian. However, a $\mathbf{k} \cdot \mathbf{p}$ model that includes the possibility of band structure anisotropy as well as  both isotropic and anisotropic third order Rashba splitting can explain the results. The isotropic third order contribution to the Rashba Hamiltonian is found to be negative, reducing the energy splitting at high $k$. The interplay of band structure, higher order Rashba effect and tuneable doping offers the opportunity to engineer not only the size of the spin-orbit splitting but also its direction. 
\end{abstract}

\maketitle

The spin-orbit interaction in solids is the basis for many fascinating phenomena, such as the spin Hall effect \cite{Hirsch:1999}, quantum spin Hall effect \cite{Kane:2005b,Bernevig:2006b,Konig:2007} and the quantum anomalous Hall effect \cite{Yu:2010}. It is also of considerable importance in conceptual spintronics devices, for instance in the Datta Das spin field effect transistor \cite{Datta:1990,Koga:2002}. An important starting point for the description of spin-orbit effects and the lifting of degeneracies in a (two-dimensional) free electron gas has been presented by Rashba and Bychkov \cite{Bychkov:1984} and this has been used to describe strong spin-orbit splittings in a variety of surface electronic states such as Au(111) \cite{Lashell:1996}, Bi \cite{Koroteev:2004}, Li/Mo(110), Li/W(110) \cite{Rotenberg:1999}, Bi/Ag(111) \cite{Ast:2007} and others. More recently, it has been found that the bulk-derived two-dimensional electron gases (2DEGs) near the surfaces of the topological insulator Bi$_2$Se$_3$ can also show a strong Rashba-type splitting and that the strength of the splitting can even be tuned by the filling of the state \cite{King:2011,Bianchi:2011,Benia:2011,Benia:2013}. Indeed, even the dispersion of the topologically protected surface states on these materials can be viewed in the context of anisotropic Rashba-type splitting \cite{Fu:2009,Frantzeskakis:2011}. For several systems, it has been found that the simple Rashba effect which is linear in $k$ is insufficient to describe the spin-orbit splitting and further correction terms such as a cubic anisotropy \cite{Fu:2009,Frantzeskakis:2011} and a cubic isotropic correction were introduced \cite{Vajna:2012}.

Here we show that strongly Rashba-split 2DEGs with intriguing new properties can be created by surface-doping of the topological insulator Bi$_2$Te$_2$Se. The band splitting in these 2DEGs is anisotropic to a degree that the bands are strongly spin-split in one direction and almost degenerate in another, possibly paving the way towards new anisotropic transport phenomena. At first glance, the observed anisotropy appears to be at odds with the spin-orbit splitting  predicted for the $C_{3v}$ symmetry of the material's (111) surface and the findings can only be reconciled with the results of a model Hamiltonian when both higher order spin-orbit splitting and pronounced band structure effects are taken into account. 

Bi$_2$Te$_2$Se crystals were grown by methods described elsewhere \cite{Mi:2013}. In the un-doped case, the bulk carrier density is sufficiently low for the observed conductance to be dominated by surface states \cite{Barreto:2014}. The samples were cleaved along the (111) surface and doped by rubidium adsorption under ultra high vacuum conditions at $\approx$ 10~K. Angle-resolved photoemission spectroscopy (ARPES) experiments were performed at the I05 beamline of the Diamond Light Source. The photon energy for the data shown here was 21~eV, the sample temperature $\approx$ 10~K and the energy and angular resolution better then 10~meV and 0.2$^{\circ}$, respectively. 

\begin{figure*}
\includegraphics[width=1.0\textwidth]{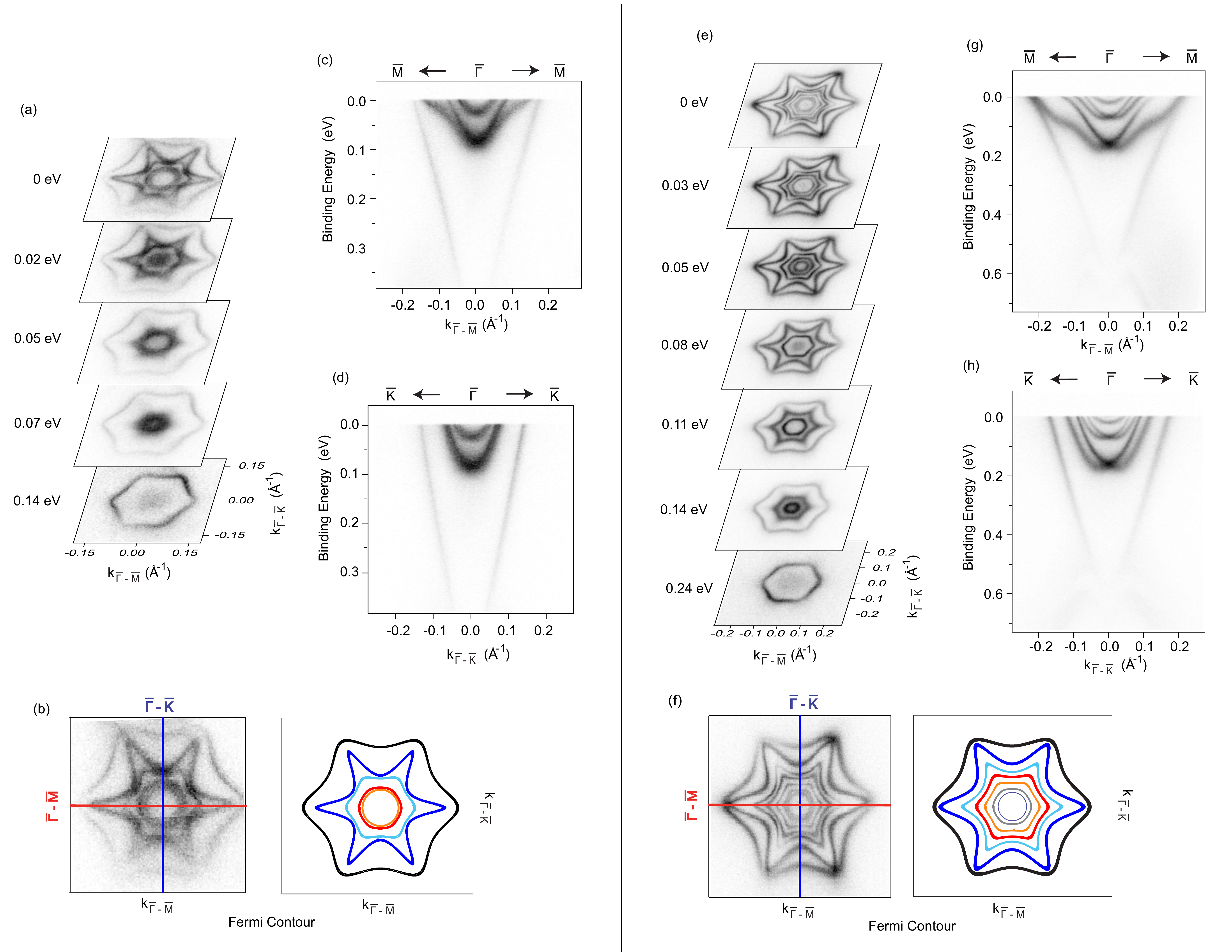}\\
\caption{(Color online) Electronic structure of electron doped Bi$_2$Te$_2$Se for the low doping case (a-d) and for the high doping case (e-h) as obtained by ARPES. (a)(e) Constant energy surfaces at different binding energies. (b)(f) On the left: Fermi contour of electron doped Bi$_2$Te$_2$Se. On the right: schematic representation of the Fermi contours. In black the topological state, in red(orange) the first 2DEG, in blue(light blue) the second 2DEG, in grey the third 2DEG. (c),(g) and (d),(h) Dispersion along the $\bar{\Gamma}\bar{M}$ and  $\bar{\Gamma}\bar{K}$ directions, respectively. }
  \label{fig:1}
\end{figure*}

ARPES spectra from two cases of electron-doping induced 2DEGs on Bi$_2$Te$_2$Se are shown in Fig. \ref{fig:1}. In the low-doping case (Fig. \ref{fig:1}(a-d)), two 2DEGs are induced in addition to the topological surface state. These give rise to complex constant energy contours in Fig. \ref{fig:1}(a) inside the hexagonal contour of the strongly warped topological surface state. For convenience we numerate the 2DEGs according to their energy minimum, with the ``first 2DEG'' being the one dispersing to the highest binding energy. In this case the second 2DEG is barely occupied, the spin-orbit splitting is observable but very small, and the constant energy contours are concentric circles. These observations are entirely consistent with the simple linear Rashba model that predicts and energy splitting of $\Delta E = 2 \alpha k$ with $\alpha$ being the so-called Rashba parameter \cite{Bychkov:1984}. For a parabolic band with effective mass $m^{\ast}$, the corresponding splitting in $k$ is $2 \alpha m^{\ast} / \hbar^2$ and this yields the concentric constant energy contours.

The first 2DEG, on the other hand, shows a pronounced spin-splitting with a strong anisotropy between $\bar{\Gamma}\bar{M}$ and  $\bar{\Gamma}\bar{K}$ (see Fig. \ref{fig:1}(c,d)). Indeed, the anisotropy is so strong that the states remain almost degenerate along $\bar{\Gamma}\bar{K}$ while their separation at the Fermi energy is $\approx 0.06$~\AA$^{-1}$ along $\bar{\Gamma}\bar{M}$. This leads to the curious Fermi contour of a hexagon inside a star with the two touching at the sides of the hexagon (see Fig. \ref{fig:1}(b)). 

 An anisotropic spin splitting is only predicted when geometric anisotropy of the potential is taken into account  \cite{Premper:2007} or higher order Rashba terms are included in a $\mathbf{k} \cdot \mathbf{p}$ model Hamiltonian \cite{Fu:2009,Frantzeskakis:2011,Vajna:2012}. For the present case of the $C_{3v}$ point group,  the symmetry-constrained energies for the splitting of a free electron-like state  are given by \cite{Fu:2009,Frantzeskakis:2011}
\begin{equation}
E(\mathbf{k}) = E_{0} + \frac{\hbar^2 k^2}{2 m^{\ast}} \pm \sqrt{\alpha^2 k^2 + \gamma^2 k^6 \cos^2 (3\phi)},
\label{equ:rashba1}
\end{equation}
where $\alpha$ and $\gamma$ are the parameters of the $\mathbf{k} \cdot \mathbf{p}$ model and $\phi$ is the angle between the two-dimensional $\mathbf{k}$ and the $\bar{\Gamma}\bar{K}$ direction. The ``$\pm$'' sign generates the outer an inner branches. The first term in the square root is the usual linear Rashba effect, which induces the in-plane spin polarization. The second term introduces an out of plane spin component and gives rise to the constant energy contour anisotropy: it is greatest along $\bar{\Gamma}\bar{K}$ and vanishes along $\bar{\Gamma}\bar{M}$, where the surface mirror line lies.

Even a qualitative inspection of the dispersion (Fig. \ref{fig:1} (c,d)) reveals it to be in disagreement with this prediction. The splitting appears larger along $\bar{\Gamma}\bar{M}$ rather then along $\bar{\Gamma}\bar{K}$. The warping of the contours introduced by the $cos^{2}(3\phi)$ term in Eq. (\ref{equ:rashba1}) must be of opposite sign for the two branches of the 2DEG. However, in the experimental results it appears as if the warping direction is the same for both branches and only the magnitude is different.

Higher filling of the states can be achieved by stronger alkali-doping, as seen in Fig. \ref{fig:1}(e-h). Here the splitting of the second 2DEG becomes more pronounced because of the higher electric field on the surface induced by the alkali-doping \cite{King:2011} and a third 2DEG is populated. However, the basic shape of the 2DEG dispersion does not change: both branches of the first 2DEG are more strongly warped, with the outer branch barely touching the topological state. The second 2DEG now shows an anisotropic splitting, too, and this splitting is in the same qualitative disagreement with Eq. (\ref{equ:rashba1}). Moreover, inspecting the dispersion of the first 2DEG's outer branch in the $\bar{\Gamma}\bar{M}$ direction (Fig. \ref{fig:1} (g)) it becomes clear that a simple free-electron-like description of the 2DEG fails. 

A likely explanation of these observations is that the dispersion of the different 2DEG branches is a combination of Rashba splitting and anisotropic non-parabolic terms in the band structure. We construct a more complete model to describe this situation by taking into account also the symmetry-constrained anisotropy and non-parabolicity within the model Hamiltonian.
The resulting eigenvalues are given by
\begin{equation}
E(\mathbf{k}) = E_{0} + \frac{\hbar^2 k^2}{2 m^{\ast}(\mathbf{k})} \pm \sqrt{(\alpha k + \beta k^3)^2 + \gamma^2 k^6 \cos^2 3\phi},
\label{equ:rashba2}
\end{equation}
where an isotropic third order term to the splitting (with the coefficient $\beta$) is introduced. With this, the splitting has the highest complexity permitted up to the third order given the symmetry constraints of the system \cite{Vajna:2012}. 
Anisotropicity and non-parabolicity are taken into account by expanding the effective mass in the kinetic energy term as a polynomial of $\mathbf{k}$ as 
\begin{equation}
m^{\ast}(\mathbf{k})=m_{0\bar{\Gamma}\bar{K}}+poly(k)\sqrt{\sin^2(3\phi)},
\label{equ:effectivemass}
\end{equation}
where $m_{0\bar{\Gamma}\bar{K}}$ is the effective mass at $\bar{\Gamma}$ along $\bar{\Gamma}\bar{K}$ and $poly(k)$ is a polynomial of order 3.

\begin{figure}
\includegraphics[width=.49\textwidth]{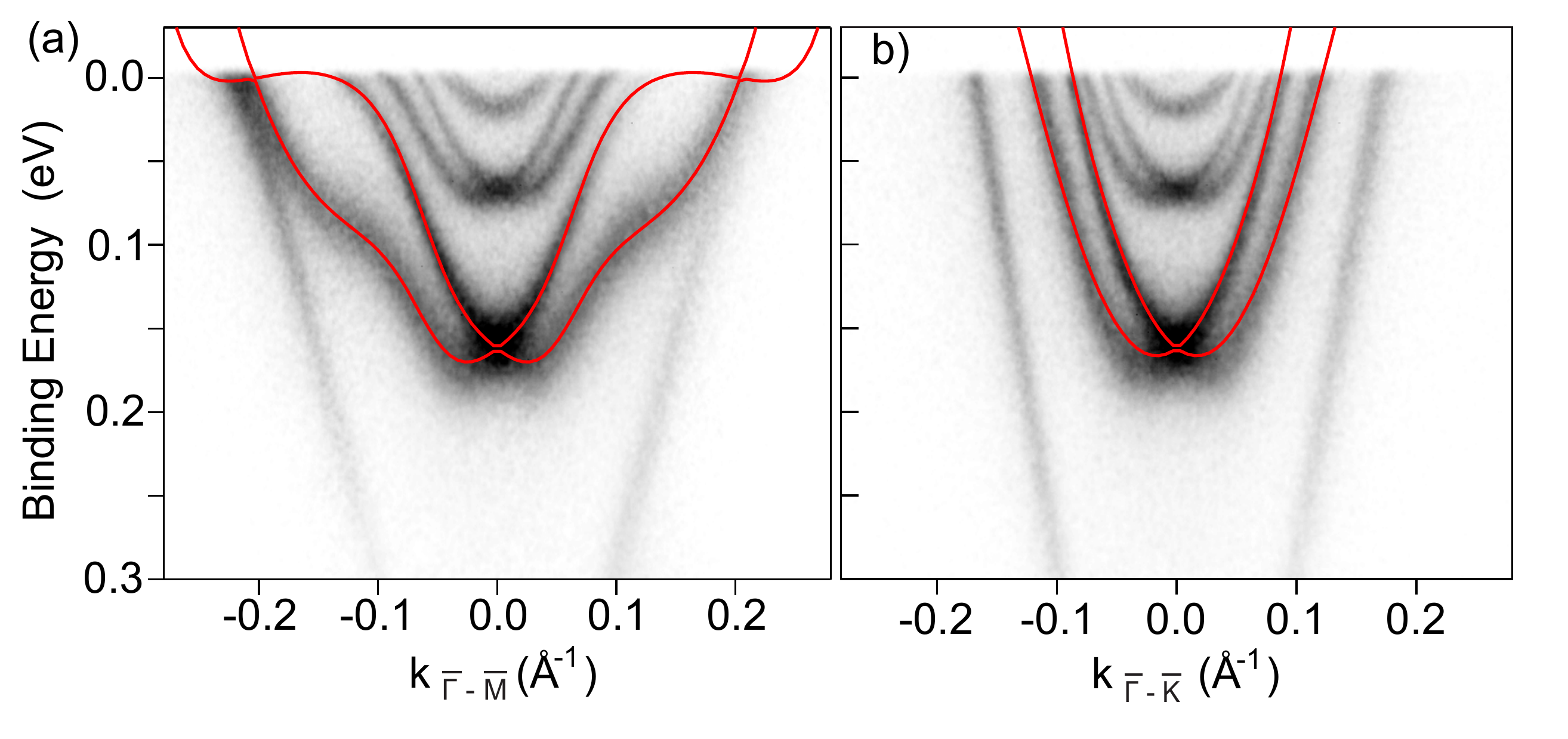}\\
\caption{(Color online) Photoemission intensity cuts of the 2DEG in the high doping case along (a) $\bar{\Gamma}\bar{M}$ and (b) $\bar{\Gamma}\bar{K}$. The red lines show the fitting of the first 2DEG branches using the model described by Eq. (\ref{equ:rashba2}).}
  \label{fig:2}
\end{figure}

This model is indeed able to fit the data well, as shown in Fig. \ref{fig:2}: red bands overlaid on the ARPES spectra represent the model used to fit the data for the two high symmetry directions in the high doping case. Note that the aim of this fit is not an accurate description of the data, we are merely interested in testing if the inclusion of higher order band structure effects can resolve the apparent inconsistency between the data and the expected anisotropy and this is clearly the case.

When fitting the band structure to the full model for the dispersion, it is difficult to obtain accurate values for the parameters because different terms in Eqs. (\ref{equ:rashba2}) and (\ref{equ:effectivemass}) can lead to equivalent dispersions. Since we are mostly interested in the Rashba part of the Hamiltonian, another approach is to consider only the actual energy splitting between the two spin-split branches. De facto this is solely described by the square root term in Eq. (\ref{equ:rashba2}), regardless of the dispersion of the state. This energy splitting is obtained from fitting energy distribution functions. This can give rise to some uncertainties very close to the bottom of the band \cite{Nechaev:2009} but these affect mostly the width of the peaks rather then their position. At higher energy and $k$, the fitting procedure is very stable.

\begin{figure}
\includegraphics[width=.49\textwidth]{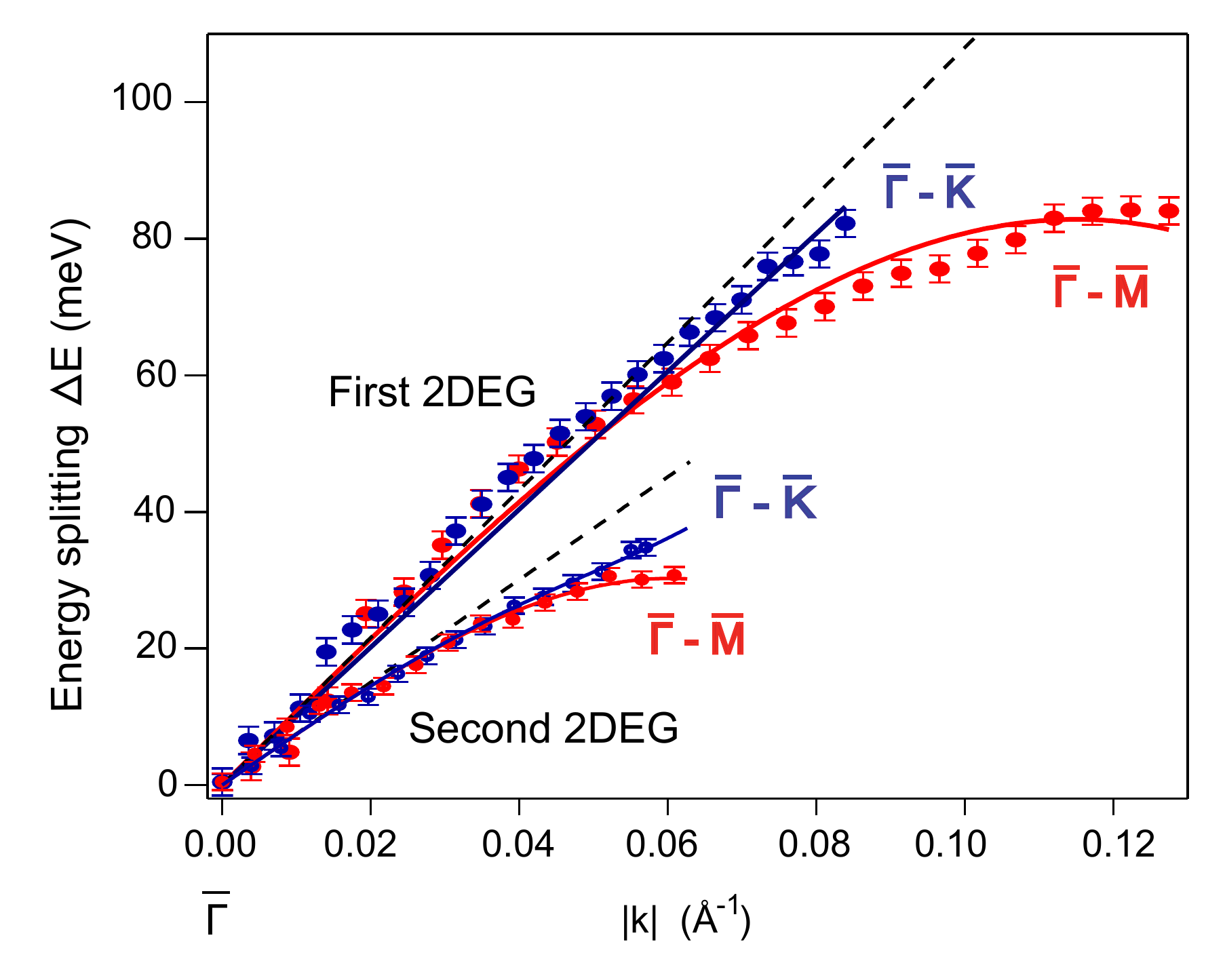}\\
\caption{(Color online) Energy splitting for the first and second 2DEGs as a function of k for the high doping case. Solid lines represent the best fit to the splitting described by Eq. (\ref{equ:rashba2}). Dashed lines represent the splitting extrapolated to first order. The observed splitting deviates from linearity at higher momentum. Both isotropic and anisotropic third order contributions are necessary for a precise description of the energy splitting.}
  \label{fig:3}
\end{figure}

The measured energy splitting of the first and second 2DEG from Fig. \ref{fig:1}(e-h) is shown in Fig. \ref{fig:3}. At low $k$, no anisotropy is found and the dashed lines show an extrapolation of the low-$k$ ($k < 0.02$~\AA$^{-1})$ data. At higher $k$, an anisotropy between the two directions is evident but it is still not very pronounced. A clear effect, however, is the deviation from a linear splitting, indicating that the isotropic third order Rashba contribution becomes significant. Interestingly, this acts to decrease the splitting of the first order Rashba effect. A similar behaviour has also been found for the Bi/Ag(111) surface alloy, extracting the dispersion from first principles calculations rather than using experimental data \cite{Vajna:2012}. 

A fit for the splitting over the entire $k$-range is shown by the solid lines in Fig. \ref{fig:3}. The fit parameters for the first (second) 2DEG are found to be $\alpha = $0.54 $\pm$~0.01 (0.37 $\pm$~0.01)~eV\AA,  $\beta = $-13.6 $\pm$~0.02 (-34 $\pm$~1)~eV\AA$^3$ and $\gamma = $34 $\pm$~3 (45 $\pm$~3)~eV\AA$^3$. The negative $\beta$ describes the reduction of the first order Rashba splitting at higher $k$. This introduces a degeneracy point for the two branches around $k = 0.2$~\AA$^{-1}$ visible in Fig. \ref{fig:2}. This crossing is likely to be avoided once the $\mathbf{k} \cdot \mathbf{p}$ Rashba Hamiltonian is expanded to even higher order because of the big $k$ value. The values obtained here for the Rashba parameter are similar to those found for the theoretical dispersion of Bi/Ag(111) \cite{Vajna:2012}. 
The energy splitting anisotropy is dictated by the $\gamma$ term that here correctly enhances the splitting along $\bar{\Gamma}\bar{K}$.

\begin{figure}
\includegraphics[width=.49\textwidth]{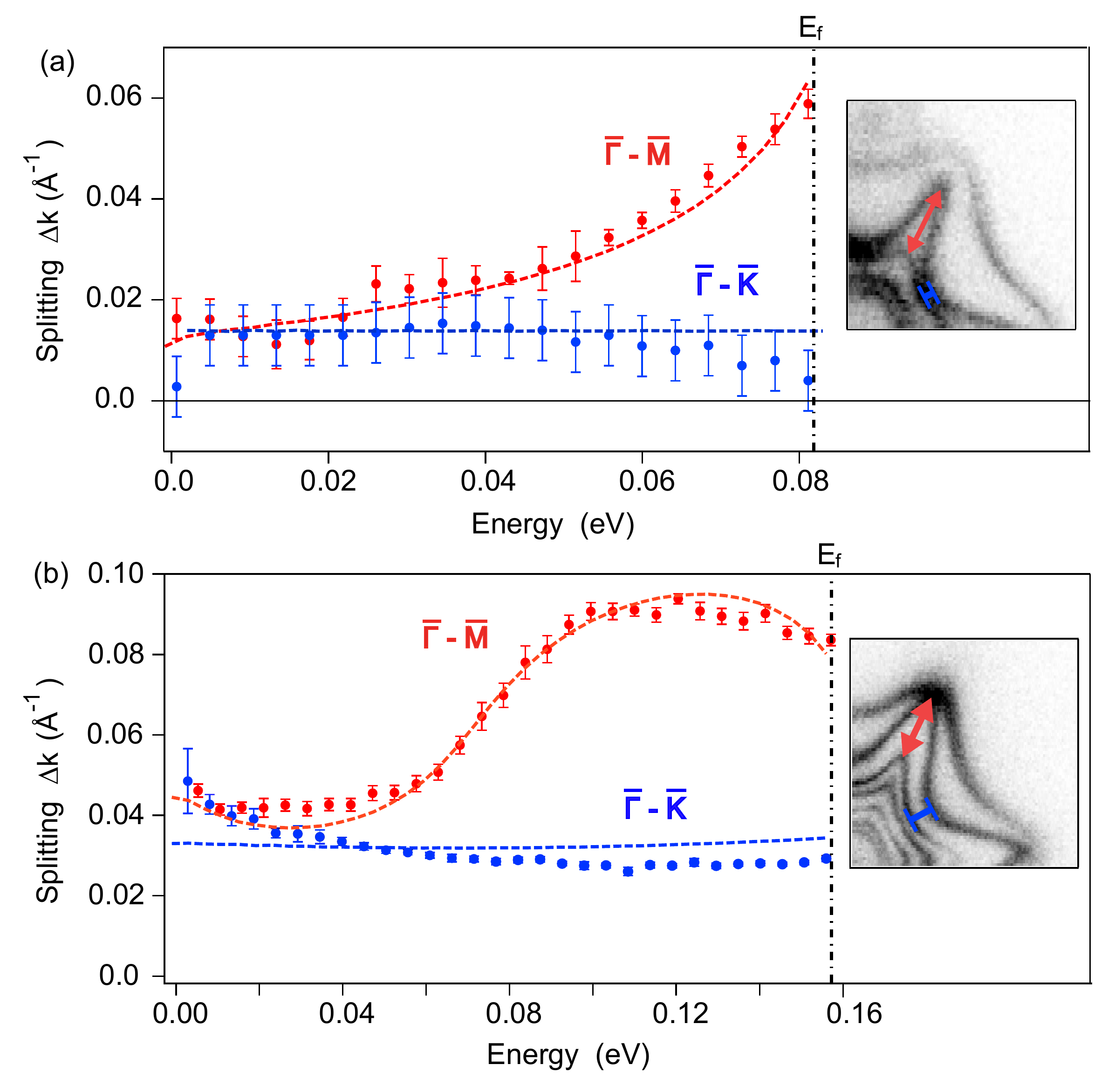}\\
\caption{(Color online) $k$ splitting of the first 2DEGs as a function of energy for the low doping (a) and high doping (b) case. The zero energy is set to be at the degeneracy point of the two branches. The symbols represent the splitting's magnitude. Dashed lines represent the $k$ splitting extracted from our model. In both cases, the combination of anisotropic band structure and Rashba spin splitting well describes the experimental results. Insets: Section of the Fermi countour that shows the anisotropy of the spin splitting at the Fermi level.}
  \label{fig:4}
\end{figure}

While the anisotropy does not look particularly pronounced when plotted as a function of $k$, the combination of band structure effects and Rashba splitting leads to a remarkable effect when plotted as a function of energy. This is visible in Fig. \ref{fig:4} for the first 2DEGs in both the low doping and high doping case. This is of extreme importance for transport experiments as it is the $k$ splitting near the Fermi energy and not the energy splitting that is actually important. The effect is biggest for the first 2DEG for which the splitting is nearly constant along $\bar{\Gamma}\bar{K}$, as expected for a Rasbha-split parabolic state, but strongly increasing in $\bar{\Gamma}\bar{M}$. This is related to the non-parabolicity and anisotropy of the band: Along $\bar{\Gamma}\bar{M}$, the dispersion deviates from the parabolic behaviour (see Fig. \ref{fig:1}(c,g)) and the outer branch assumes a negative effective mass. The inner branch, on the other hand, maintains its positive effective mass. This introduces the strong enhancement of the spin splitting even when the Rashba coupling term is actually very small as in the low doping case. For the higher doping case we can look even further up in the structure of the 2DEG: below 0.09 eV binding energy the outer branch re-assumes a positive curvature while the inner branch starts to bend downward and therefore the splitting slowly decreases.
Even though the situation is qualitatively similar for both doping levels, the doping still has a big effect on the Fermi contour texture since the splitting along $\bar{\Gamma}\bar{K}$ can be driven to almost zero for lower doping while the splitting along $\bar{\Gamma}\bar{M}$ remains relatively large. 

A qualitative understanding of the anisotropic splitting in the 2DEG can be derived from the connectivity of the surface states and the bulk states in a topological insulator \cite{Wray:2011,Bahramy:2012}. Time-reversal and crystal symmetry impose constraints on the splitting of the surface state and the 2DEG. These states can be split anywhere in the surface Brillouin zone except for the so-called time-reversal invariant momenta $\bar{\Gamma}$ and $\bar{M}$ where the splitting has to vanish. At $\bar{\Gamma}$ this is clearly observed for all the states and doping levels and it is also predicted by the Rashba-type Hamiltonian of highest complexity (see Eq. (\ref{equ:rashba2})). At $\bar{M}$, a simple free-electron-like Rashba model would not be able to guarantee the required degeneracy (even though the isotropic third-order Rashba term could lead to a coincidental degeneracy). In the experiment, the states at $\bar{M}$ are unoccupied and thus not observable by ARPES. Still, the crystal symmetry dictates that the topological surface state needs to be connected with a state of opposite spin at $\bar{M}$ to achieve degeneracy. This has been discussed for Bi$_2$Se$_2$ \cite{Wray:2011} and it has been shown that the topological state connects to the outer branch of the first 2DEG while the inner branch of this 2DEG connects to the outer branch of the second 2DEG \cite{Bahramy:2012}. 

These considerations might well lead to the anisotropy of the splitting observed here: Along $\bar{\Gamma}\bar{M}$ the splitting between the surface state and the outer branch of the 2DEG eventually has to vanish at $\bar{M}$ while there is no such requirement at $\bar{K}$. Therefore the outer branch of the first 2DEG has to follow the warping of the topological state even though the same third order anisotropic Rashba term that is responsible for the warping of the surface state would lead to a warping that goes into the opposite direction. We are thus faced with a situation where the symmetry requirements of the spin splitting influence the higher order terms in the non-spin dependent part of the dispersion. 

In conclusion, we have shown Bi$_2$Te$_2$Se to host 2DEGs with a strong and anisotropic  spin-splitting.  Superficially, the anisotropy is at odds with the symmetry predicted from a Rashba Hamiltonian including third order anisotropic corrections but the entire dispersion can be described correctly if anisotropic higher order terms are introduced in the spin-independent part of the dispersion. The anisotropy of the splitting can be very large in $k$, up to a point where it vanishes in one crystallographic direction while being simultaneously large in another. In combination with the band structure effects, it is even possible to find energies where the effective mass of the electrons changes sign between the two split branches in a given direction. One should expect interesting and tuneable spin-dependent transport phenomena in this and similar systems. In fact, transport effects in the presence of both Rashba and Dresselhaus splitting have been discussed in some detail \cite{Schliemann:2003,Schliemann:2003b,Ganichev:2004} but higher-order Rashba contributions have so far not been included in these considerations. The present system is now also becoming accessible in transport measurements, as recently shown using quantum oscillations  \cite{Cao:ARXIV}.

We gratefully acknowledge financial support from the VILLUM foundation, The Danish Council for Independent Research, the Danish National Research Foundation (DNRF93) and the Lundbeck Foundation.


\end{document}